\documentclass[aps,prb,twocolumn,showpacs,floatfix,superscriptaddress]{revtex4-1}

\usepackage{graphicx,color}
\usepackage{amsfonts}
\usepackage[figuresright]{rotating}  
\usepackage{amssymb}
\usepackage{amsmath}
\usepackage{mathtools}
\usepackage{psfrag}
\usepackage{subfigure}
\usepackage{multirow}
\usepackage{tabularx}
\usepackage{textcomp}
\usepackage{units}
\usepackage{hyperref}
\hypersetup{
 pdfnewwindow=true, colorlinks=true,
 linkcolor=blue, anchorcolor=blue,
 citecolor=blue, filecolor=blue,
 menucolor=blue, urlcolor=blue}

\def\beq{\begin{eqnarray}}
\def\eeq{\end{eqnarray}}

\renewcommand{\v}[1]{\ensuremath{\mathbf{#1}}} 
 
\let\baraccent=\= 
\renewcommand{\=}[1]{\stackrel{#1}{=}} 
	
\makeatletter

\begin{document}
\title{Intrinsic surface dipole in topological insulators}
\author{Benjamin M.\ \surname{Fregoso}}

\affiliation{Department of Physics, University of California, Berkeley,
California, 94720, USA}

\author{Sinisa\ \surname{Coh}}
\affiliation{Department of Physics, University of California, Berkeley,
California, 94720, USA} 
\affiliation{Materials Sciences Division, Lawrence Berkeley National Laboratory, 
Berkeley, California 94720, USA}

\begin{abstract}
We calculate the local density of states of two prototypical
topological insulators (Bi$_2$Se$_3$ and Bi$_2$Te$_2$Se) as a function
of distance from the surface within density functional theory. We find that, in the absence of
disorder or doping, there is a 2~nm thick surface dipole the origin of which
is the occupation of the topological surface states above the
Dirac point.  As a consequence, the bottom of the conduction band is
bent upward by about 75~meV near the surface, and there is a hump-like feature 
associated with the top of the valence band. We expect that band
bending will occur in all pristine topological insulators as long as
the Fermi level does not cross the Dirac point. Our results show that 
topological insulators are intrinsic Schottky barrier solar cells.
\end{abstract}

\pacs{73.20.At,73.30.+y,71.20.-b,77.90.+k,73.50.-h, 73.61.-r}
\maketitle

\section{Introduction}
\label{sec:intro}
Semiconductor interfaces are of fundamental importance 
in electronic device operation. At the heart of this technology are  
heterojunctions made of n-doped and p-doped semiconductors\cite{Monch2010}.
The charge carriers through $pn$-junctions can be accurately controlled 
via gate voltages, giving diodes and transistors its useful properties.
It is well known that near the interface region there is a permanent electric field, which 
bends the conduction and valence levels. This region is referred to as space-charge region or  
interface dipole (\textit{surface dipole} if the interface is with vacuum). 
This band-bending (BB) effect is also of fundamental importance in silicon-based solar 
cell technology, where the electric field separates photo-exited electrons and holes.

Band bending in semiconductors can be determined from the I-V 
characteristics of the junction, from surface photo-voltage spectroscopy\cite{Kronik1999}, 
surface potential microscopy or from Angle-Resolved Photoemission Spectroscopy (ARPES).
Many factors contribute to the BB measured by any of these techniques.
Among the most important are the difference in work functions of the materials at the interface, the image charge potential 
and possible interface reconstruction. These are `intrinsic' contributions.
Occupied impurity states also contribute to BB (extrinsic contributions).
Analytical calculations of BB in semiconductor interfaces usually determine the charge
distribution self-consistently from an effective one-dimensional Poisson equation
and a single-particle Schrodinger equation. Such calculations often depend on unknown parameters, 
which are ultimately determined in experiments. 

\begin{figure}[t!]
\subfigure{\includegraphics[width=0.48\textwidth]{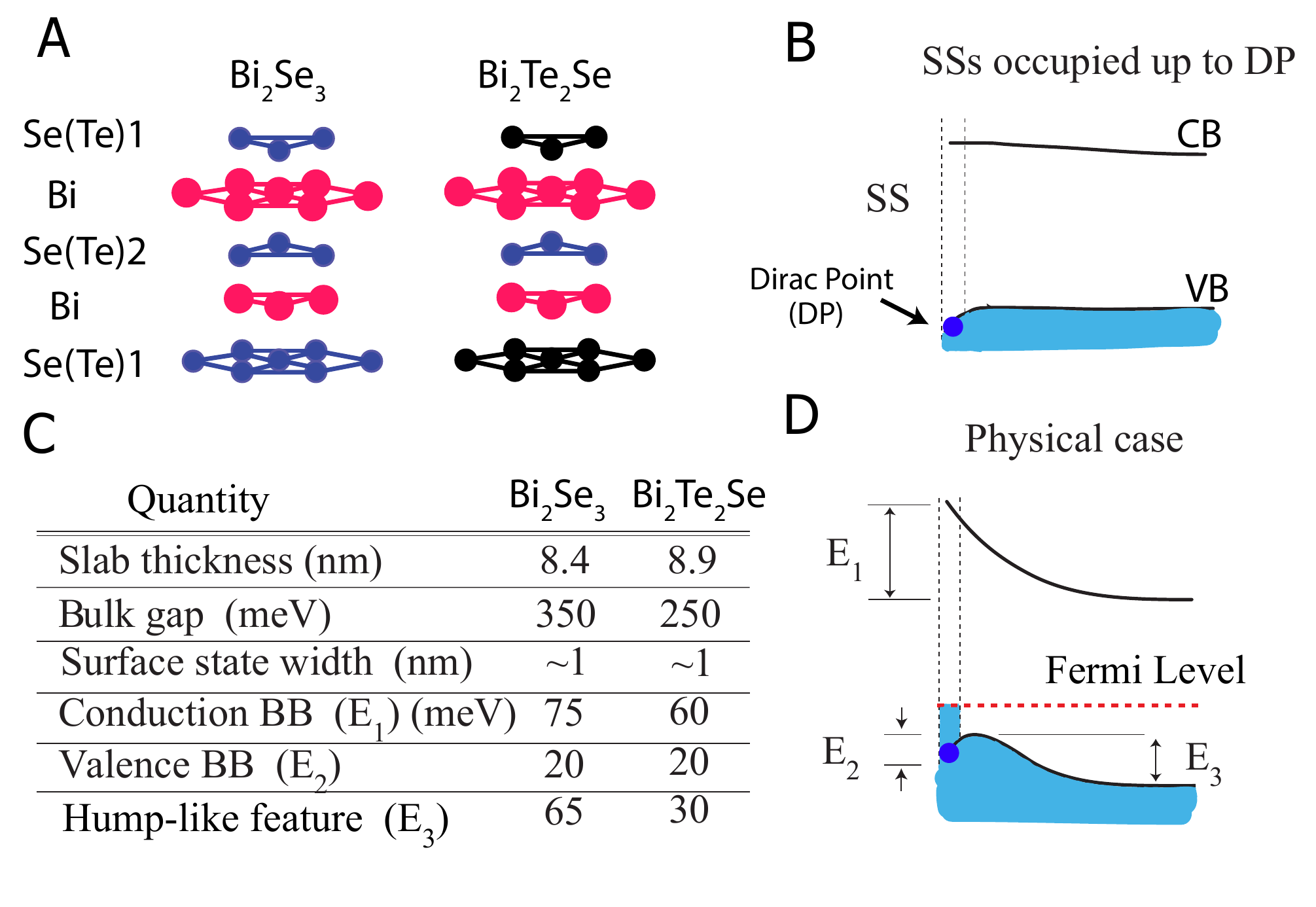}}
\caption{(Color online) (A) Hexagonal crystal structure of 
Bi$_2$Se$_3$ and Bi$_2$Te$_2$Se quintuple layers. (B) schematic diagram shows 
a cross section of the slab and the location of the valence band (VB) and conduction band (CB). 
Occupied regions are shaded. Two cases are shown: (B)
when the surface states (SSs) are occupied up to the Dirac point (DP) and (D) when they are thermally occupied. When the 
SSs are occupied up the the DP only (non-thermal case), \textit{upward} band bending (BB) is absent but
there is a small downward BB near the surface. When SSs are occupied (thermal case), 
the bands bend upward near the surface of the slab. In (C) we 
show the BB characteristics of Bi$_2$Se$_3$ and Bi$_2$Te$_2$Se. 
E$_{1,2,3}$ parametrize the magnitude of BB.}
\label{fig:table}
\end{figure}

Recently, a new type of `semiconductor' was discovered, 
the topological insulator (TI), which is an insulator in the bulk but has  metallic surface
states (SSs) protected by time reversal symmetry\cite{Roy2009,Moore2007,Fu2007,Dzero2010,Zhang2013}. 
For further details about TIs we refer the reader to the many excellent reviews of the subject such as 
Ref.~\onlinecite{Hasan2011} and references therein. At low energies, the surface states of TIs can be described 
by the Dirac equation in two-dimensions with the spin of the electron locked to its momentum. This property can be 
utilized in spintronic\cite{Yokoyama2014} applications.

Band bending in TIs surfaces has been investigated experimentally with ARPES
\cite{Hsieh2009,Bianchi2010,Analytis2010,Chen1012,Neupane2012,Edmonds2014,Frantzeskakis2015,Urazhdin2004} and 
other techniques\cite{Pettes2013,Narayan2014}.
However, the results have not been consistent, even in apparently similar conditions.
For example, BB of Bi$_2$Se$_3$ has been found to be either upward
($\sim$75~meV)\cite{Analytis2010} or downward
($\sim$100--300~meV)\cite{Hsieh2009,Bianchi2010,Chen1012,ViolBarbosa2013,Frantzeskakis2015,Urazhdin2004}.
This suggest that the dominant contribution to BB in these experiments is extrinsic, e.g., doping, and 
hence it is very sensitive to the details of the sample preparation. 
In view of these considerations, it is clear that
extrinsic and intrinsic effects play a role in actual experiments and that  
disentangling intrinsic from extrinsic factors would be very useful as a benchmark for future 
experimental and theoretical studies. On the theoretical side, the Poisson-Schrodinger 
equations\cite{King2008,Brahlek2015} and tight-binding models\cite{Galanakis2012,Bahramy2012} have been successfully  
used to model BB in TIs. However, understanding and characterizing BB in TIs interfaces has not been explored 
systematically from first principles despite the fact that it could open the door for new technological applications, e.g., in solar cell technology.

In this paper we resolve the intrinsic contribution to BB (and hence to the surface dipole) by studying 
BB in pristine TIs Bi$_2$Se$_3$ and Bi$_2$Te$_2$Se. We consider a finite slab of Bi$_2$Se$_3$ and Bi$_2$Te$_2$Se
and calculate variations of the local density of states (LDOS) 
as a function of depth using first principles density functional theory (DFT) 
with no fitting parameters. Fig.~\ref{fig:table} summarizes our main results. We find 
an upward BB of the conduction and valence bands within 2~nm below the
surface. The upward BB for Bi$_2$Se$_3$ and Bi$_2$Te$_2$Se is 
75~meV and 60~meV respectively, even without disorder or doping.  
We attribute this BB to the occupation of topological SSs above the Dirac point.
In the valence band we find a hump-like feature near the surface consisting of an  
upward bend (65~meV for Bi$_2$Se$_3$ and 
30~meV for Bi$_2$Te$_2$Se) 2~nm from the surface, followed by a 20~meV
downward bend 1~nm from the surface. 
We note that, our results do not rely on the Poisson equation to find the equilibrium 
charge distribution. Instead, the charge distribution arises
from the inclusion of electron correlations at the DFT level.
Therefore, we expect deviations from Poisson electrostatics.
In particular, we find asymmetric BB of the valence and conduction bands,
including a small downward BB of the valence band near the surface.
Finally, although we used Bi$_2$Se$_3$ and Bi$_2$Te$_2$Se as TI prototypes, 
we expect similar order of magnitude BB for the family Bi$_2$X$_3$ with X=Se,Te 
and X$_2$Te$_2$Y with X=Sb,Bi and Y=S,Se, as Y and X enter with the same oxidation state.

This paper is organized as follows. In Sec.~\ref{sec:TI_ldos} we compute the band structure and 
LDOS of Bi$_2$Se$_3$ and Bi$_2$Te$_2$Se for the physically relevant situation where thermalized 
electrons fill the SSs above the Dirac point. In Sec.~\ref{sec:SD} we compare with the LDOS
without spin-orbit coupling and with SSs occupied up to the DP and 
discuss the origin of the surface dipole. 
The appendix~\ref{app:detail_numerics} contains
additional information about the calculation.

\section{Local density of states}
\label{sec:TI_ldos}
\subsection{Crystal structure}
\label{sec:crystal_struc}
\begin{figure}[t!]
\subfigure{\includegraphics[width=0.48\textwidth]{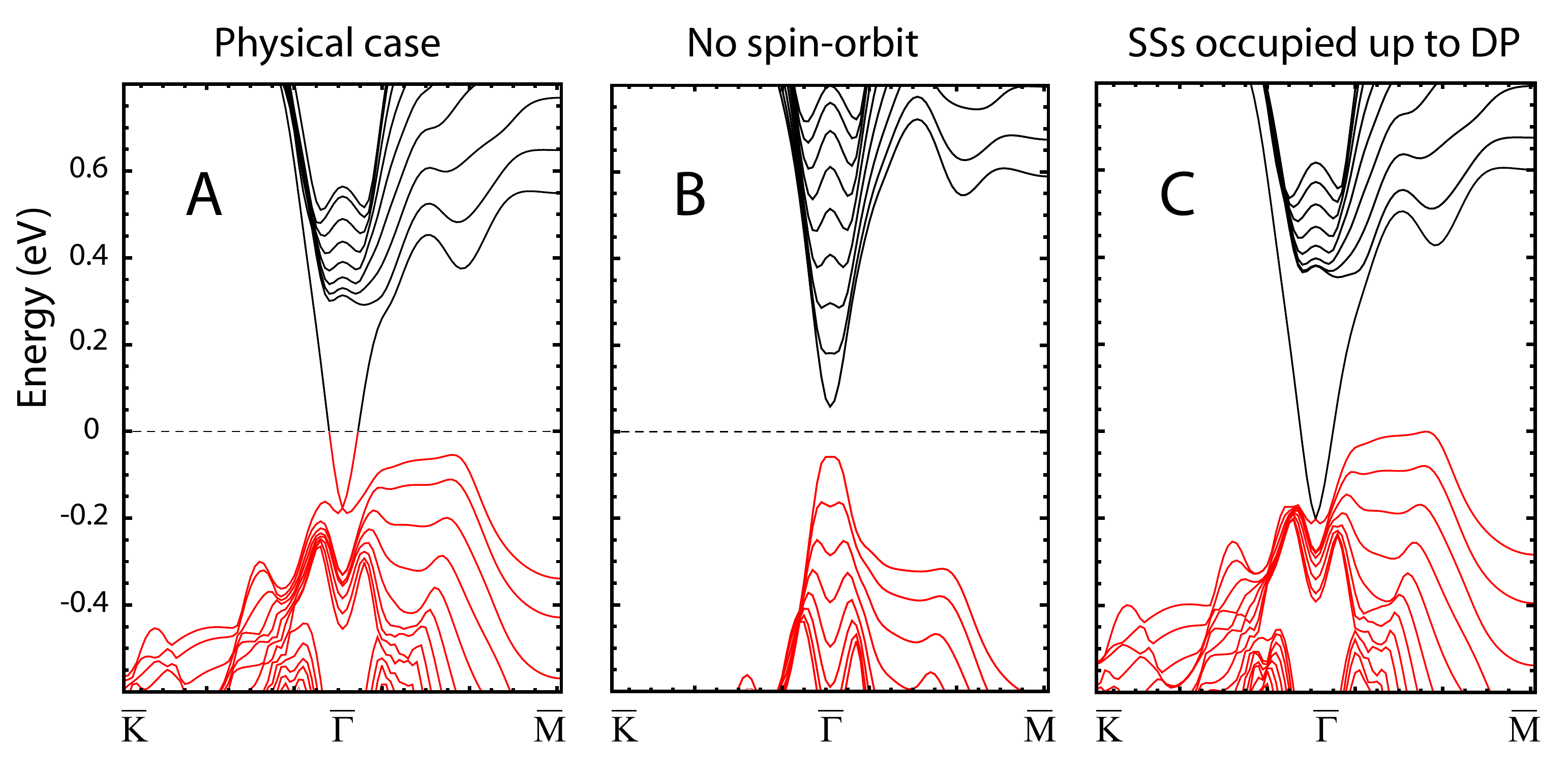}}
\caption{(Color online) Band structure of a slab of Bi$_2$Se$_3$. 
(A) corresponds to the physical case where electrons are thermalized at 300 K and 
occupy surface states (SSs) above the Dirac point (DP) which is located at the $\bar{\Gamma}$ point.  
In (B) the spin-orbit coupling has been turned off and hence there are no SSs. In (C) 
the SSs are occupied up to the DP. In all panels occupied bands are marked in red.}
\label{fig:bands_struct}
\end{figure}

Bi$_2$Se$_3$ and Bi$_2$Te$_2$Se are layered materials with
rhombohedral (R$\bar{3}$m) crystal structure\cite{Cava2013}.  The
structure can be represented as hexagonal planes stacked along the
$c$-axis containing only one type of atom,  as shown in Fig.~\ref{fig:table}A.   
In one hexagonal unit cell, the order of the stacking is {\small
  -[Se(2)-Bi-Se(1)-Bi-Se(2)]$_0$-[Se(2)-Bi-Se(1)-Bi-Se(2)]$_{1/3}$-[Se(2)-Bi-Se(1)-Bi-Se(2)]$_{2/3}$}
in Bi$_2$Se$_3$ and {\small
  -[Te-Bi-Se-Bi-Te]$_0$-[Te-Bi-Se-Bi-Te]$_{1/3}$-[Te-Bi-Se-Bi-Te]$_{2/3}$-}
in Bi$_2$Te$_2$Se.  The group of five atomic layers in square brackets
are often referred to as a quintuple layer (QL) shown in Fig.~\ref{fig:table}A.  
The subscripts indicate the fractional translation of the QL along the $c$-axis of the
hexagonal unit cell.  The QLs are weakly coupled by Van der Waals
forces, forming natural cleavage planes with negligible 	
surface reconstruction. In what follows we present results for 
Bi$_2$Se$_3$ only, the results for Bi$_2$Te$_2$Se are 
similar and summarized in Fig.~\ref{fig:table}.

\subsection{Band structure and local density of states}
\label{sec:band_stru_ldos}

Here we provide some details of our first-principles calculation of
the LDOS (more information is in the
appendix~\ref{app:detail_numerics}).  We used
Quantum-ESPRESSO\cite{Giannozzi2009} and Wannier90\cite{Mostofi2008}
to compute the band structures and LDOS of Bi$_2$Se$_3$.  Our periodic computational unit cell consists of 9
QLs of Bi$_2$Se$_3$ (total width of 8.4 nm) 
with a 1 nm of vacuum layer between periodic
images. Therefore, each computational cell consists of 45 atoms in
total, at which the LDOS is calculated. Bi$_2$Se$_3$ slabs are
terminated with Se atoms on both sides, as found in experiments\cite{Cava2013}.
Other terminations are possible but generally lead to reconstructed surfaces
which are absent in our case.

\begin{figure}[t!]
\subfigure{\includegraphics[width=0.48\textwidth]{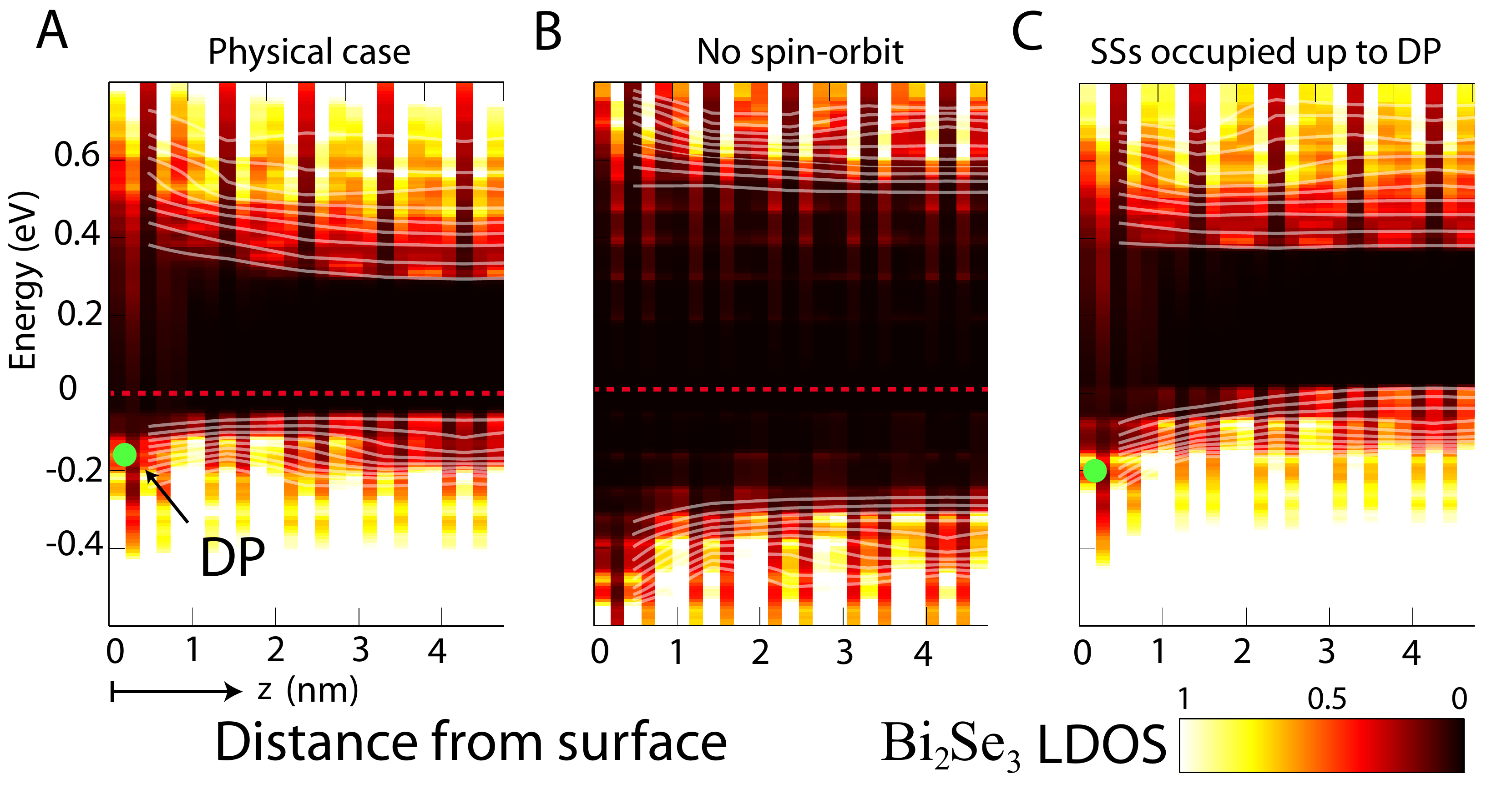}}
\caption{(Color online) (A) Local density of states (LDOS) across the slab
corresponding to the physical case in which electrons are thermalized at 300~K and 
occupy SSs above the Dirac point (DP). (B) LDOS with no spin-orbit interaction and (C) 
occupying surface states (SSs) up to the DP. About half of the slab is shown in each case. 
Dashed lines indicate the Fermi level when appropriate, and the green dot indicates the position of 
the DP.  The color scale used in panels indicates the LDOS in arbitrary 
units (largest LDOS corresponds to white color).  The energy
resolution is 3~meV.  The white lines are contours of equal intensity 
of LDOS. The contours where smoothed by Gaussian averaging ($\sigma=36$ meV) over energy and in each 
quintuple layer (QL).}
\label{fig:ldos_z}
\end{figure}

To visualize BB we compute
the LDOS across the entire slab and then note its variations in intensity 
as a function of depth. To obtain the LDOS with a required energy resolution we
use the Wannier interpolation\cite{Yates2007} method. This method
consists of first computing Kohn-Sham orbitals $\psi_{n\v{k}}$ and
energies $\epsilon_{n\v{k}}$ on a coarsely sampled Brillouin zone (we
used a uniform 9$\times$9 mesh) for electronic bands of interest. 
In Fig.~\ref{fig:bands_struct}A we show the band structure 
of Bi$_2$Se$_3$.  The Dirac point (DP) is at the $\bar{\Gamma}$ point of the Brillouin zone 
180 meV below the Fermi level. Note that the SSs (above the DP) are occupied.
and there is a bulk gap of $\sim350$~meV consistent with 
previous reports\cite{Black1957,Xia2009,Xu,Lovett1977}.
We then use these orbitals to construct localized Wannier
functions\cite{Marzari2012} $|{\bf R} m i \rangle$ centered at lattice
vector $\bf R$, with both $m=s,p$ like atomic character for
all 45 layers in the slab (index $i$ goes from 1 to 45). Since the
Hamiltonian is exponentially localized in the Wannier basis we can use this
basis to interpolate electronic orbitals and energies on a very dense
k-mesh (3000$\times$3000) with a negligible additional computational cost.
Finally, we obtain the LDOS from the histogram of
the interpolated electronic energies $\epsilon_{n\v{k}}$,

\begin{align}
N(E, i) = \sum_{n \v{k} m} 
|\langle {\bf 0} m i |\psi_{n\v{k}}\rangle|^2 \delta(E - \epsilon_{n\v{k}}).
\label{eqn:ldos}
\end{align}
Here the sum is calculated over all electronic states (band $n$ and momentum
$\v{k}$), and over all Wannier functions ($m=s,p$), so that
the local density of states $N(E,i)$ depends only on atomic layer
($i$) and energy ($E$). The LDOS (shown in Fig.~\ref{fig:ldos_z}A) was projected at each of the 45
atomic layers of our supercell. The horizontal axes 
indicates the approximate distance from the left surface.  
The matrix element appearing in the
sum is independent of lattice vector ${\bf R}$, since the Bloch
functions are cell-periodic. Therefore we can sum only over ${\bf R}={\bf 0}$ for simplicity.

Fig.~\ref{fig:ldos_z}A shows the LDOS across 
a slab of Bi$_2$Se$_3$ in the physically relevant case with thermalized electrons
occupying the SSs. For comparison we also include the case without 
spin-orbit coupling (B) and with occupied SSs up to the DP only in (C). 
In this section we discuss the physically relevant case Fig.~\ref{fig:ldos_z}A.
The discussion of cases without spin orbit coupling and with partial filling 
of SSs is given in Sec.~\ref{sec:soc} and  Sec.~\ref{sec:popu_SS}.

Near the surface there is a non-zero spectral weight 
extending roughly $d \sim 1$~nm into the bulk,
which originates from the topological SSs. Interestingly,
one can follow the edges of the valence and conduction bands by 
observing how the spectral weight \textit{intensity} varies across the
slab. To quantify BB we follow the following procedure. We define BB by  
the maximal deviation of the LDOS isocontours (see white lines in Fig.~\ref{fig:ldos_z}A)
between the first and middle QLs in a given energy window (200~meV below the valence band
maximum or above the conduction band minimum). With this definition,
the edge of the conduction band bends upward by $\sim 75$~meV within 
2~nm of the surface, Fig.~\ref{fig:ldos_z}A. On the other hand, the 
valence band has a hump-like feature near the surface
consisting of a 65~meV upward bend 2~nm from the surface followed by a 20~meV downward bend 1~nm from the
surface.

\section{Surface dipole}
\label{sec:SD}

In this Section we study the origin of BB in Bi$_2$Se$_3$ 
and show that it can be modeled as a macroscopic surface dipole\cite{Monch2010}.
We will study both the effect of the spin-orbit interaction and the 
population of the surface states.

\subsection{Spin-obit interaction}
\label{sec:soc}
Turning off the spin-orbit interaction in our calculation reduces
the TI to a non-topological insulator, with no topological SSs.
This can be seen by the lack of edge states in Fig.~\ref{fig:bands_struct}B.
We expect that with no topological SS, the charge distribution across the
slab will be affected. This
is indeed what we find, see Fig.~\ref{fig:ldos_z}B.  Without
spin-orbit interaction, the conduction band is nearly flat.  The
maximum bending is less than 20~meV.  
Therefore, we can assign the origin of the
upward band bending of the conduction band to the presence of the
topological SSs.
We focus on the occupation of these states in more detail in the next subsection.

Focusing now on the valence band, we find that even without spin-orbit
coupling there is a 75~meV downward BB and a much smaller
hump-like feature. Therefore, we again assign the origin of the
upward BB to the presence of topological SSs.

\subsection{Population of the surface states}
\label{sec:popu_SS}
While the presence of topological SSs is essential for BB
in pristine TI slabs, here we discuss the importance of their filling. 
As discussed in Sec.~\ref{sec:TI_ldos}, the DP of the
Bi$_2$Se$_3$ slab is below the Fermi level,
and hence some of the SSs which are above the DP are populated, Fig.~\ref{fig:bands_struct}A.
Since the whole slab is neutral, this excess charge on the surface
must be compensated by a partial depopulation of some of the bulk-like
states. 

Fig.~\ref{fig:ldos_z}C shows the LDOS for a Bi$_2$Se$_3$ slab with SSs 
occupied up to the Dirac point only and unoccupied above it, see also Fig.~\ref{fig:bands_struct}C. 
This can be accomplished by setting the population of 
the first 252 valence bands at all k-points to unity. As can be seen, most of the upward BB 
features disappear. The conduction BB is only 20~meV across
the slab.  For the valence band, maximal upward bending (within 200~meV of the valence
band maximum) is only 15~meV.  
However, a downward bend of the valence band near the
surface is still present. The downward bend is 90~meV, 
even larger than in the thermalized case.

Therefore, to conclude this and the previous subsection, we find that
upward bending of the conduction and valence bands is present only
when topological SSs are present (Sec.~\ref{sec:crystal_struc}) and
when they are occupied. 

\subsection{Magnitude of the surface dipole}
As a consistency check we compare the surface charge density estimated
in two ways. First, we calculate the phase space occupied by the SSs
from the DP up to the Fermi level in our thermalized calculation. This
estimate gives us a surface charge density $n_{\rm 2D} \sim 0.7\times
10^{13}$~cm$^{-2}$. Second, we estimate the surface charge density by
modeling the region near the surface as an electric double layer
separated by about $L=2$~nm with opposite charges.  One layer is
negative due to the occupation of the SS and the other is positive
from the depletion of electrons in the bulk-like region. In 
Fig.~\ref{fig:ldos_z}b, we see that the potential discontinuity at the surface is
$\Delta V\sim 75 $~meV and from $\Delta V = D_{\rm SS}/\epsilon$
(with $\epsilon \sim 100\epsilon_0$ the permittivity of
Bi$_2$Se$_3$~\cite{Madelung1998}) we obtain $D_{\rm SS}\sim 6\times
10^{-11}$~C/m.  Using $D_{\rm SS}=e n_{\rm 2D} L$ we find a surface
density $n_{\rm 2D}\sim 2\times 10^{13}$~cm$^{-2}$ in a reasonable
agreement with the first estimate, thus confirming that the occupation
of SS above the DP is responsible for the observed band bending.

\section{Conclusion}
\label{sec:conc}

We have shown that free surfaces of pristine Bi$_2$Se$_3$ and
Bi$_2$Te$_2$Se with no doping or disorder have an intrinsic surface dipole.  
Bi$_2$Se$_3$ and Bi$_2$Te$_2$Se grown without any post-processing 
tend to be doped and disordered and appear to be good bulk
conductors\cite{Ren2010,Bianchi2010,Neupane2012,Yang2012,Marcinkova2013,Brahlek2015,Arita2014}.
However, even in this case the upward contribution to band bending computed in this work is present. 
We note that Ref.~\cite{Analytis2010} finds $\sim$75 meV upward band bending in Bi$_2$Se$_3$ which is 
similar to pristine Bi$_2$Se$_3$. However, more analysis is required to 
attribute this band bending solely to intrinsic effects.
As samples become cleaner\cite{Hoefera2014} only the intrinsic component computed here will remain. 
We comment that, we have considered the case where the DP lies below the Fermi level.
We leave for future work the discussion of the case where the Fermi level crosses at or below 
the DP. 

Importantly, the intrinsic band bending found in this work means that TIs generate up to  
$\sim$ 75 meV of surface photvoltage\cite{Kronik1999} upon illumination. 
The same principle applies in the operation of Schottky barrier solar cells, where the interface dipole  
is created at a metal-semiconductor interface. Therefore, TIs could operate as \textit{intrinsic} Schottky barrier solar cells
with an estimated maximum efficiency of $\sim$ 7\% according to the Shockley-Queisser\cite{Monch2010} criterion.

After completion of this work we became aware of Ref.~\onlinecite{Rakyta} which
considers BB in Bi$_2$Se$_3$.

\section{Acknowledgments}
We thank J. Moore, J. Analytis, M. L. Cohen, C. Salazar, C. Ojeda-Aristizabal, and A. Drieschova for useful discussions.
Support was provided by Conacyt and NSF DMR-1206513, and Computer resources by NERSC 
under Contract No. DE-AC02-05CH11231.

\appendix
\section{Numerical details}
\label{app:detail_numerics}

We use density functional theory (DFT) as implemented in the
Quantum-ESPRESSO\cite{Giannozzi2009} computer package with the
generalized gradient approximation\cite{Perdew1997} to the
exchange-correlation energy functional. Convergence of the energy in
the self-consistent iterations was verified to better than
10$^{-8}$~Ry and we used a 100~Ry plane wave kinetic energy cutoff.  A
k-point mesh of 9$\times$9$\times$1 (9$\times$9$\times$9) was used
in slab (bulk) calculations.  We performed a fully-relativistic
calculation with relativistic effects included in the
pseudopotentials.

We first fully relax the lattice constants and the internal
coordinates in bulk compounds.  For Bi$_2$Se$_3$ we obtained the DFT
optimized parameters $a=4.080$~{\AA}, $c=28.198$~{\AA} and for
Bi$_2$Te$_2$Se $a=4.247$~{\AA}, $c=29.632$~{\AA}. Following a bulk
structural relaxation we performed the relaxation of the internal
coordinates in the slab geometry. We checked that the LDOS results
did not change if we used 6~QL instead of 9~QL or if we used 20~{\AA}
vacuum instead of 10~{\AA}. We then constructed maximally 
localized Wannier functions\cite{Marzari2012}
and performed the Wannier interpolation\cite{Yates2007} using the
Wannier90\cite{Mostofi2008} computer package.  For initial projections
we used $s$ and $p$-like atom-centered orbitals for both Bi and Se
(Te) atoms.

In the non-thermal calculation discussed in section \ref{sec:popu_SS},
the occupations of the lowest $N$ bands at each k-point are fixed to
one, where $N$ is the total pseudocharge of our unit-cell
($N$=252). Filling up the lowest $N$ bands in a topological insulator
corresponds to occupying states up to the Dirac point.

%

\end{document}